\documentclass{article}
\usepackage{spconf,amsmath,graphicx}
\usepackage{enumitem}
\usepackage{multirow}
\usepackage{subfigure}
\setlist{nosep, leftmargin=14pt}

\usepackage{mwe} 
\usepackage{listings}
\usepackage{algorithm}
\usepackage{algorithmicx}
\usepackage[hidelinks]{hyperref}
\hypersetup{
  colorlinks   = true, 
  urlcolor     = magenta, 
  linkcolor    = red, 
  citecolor   = green 
}

\title{MAG: A SIMPLE LEARNING-BASED PATIENT-LEVEL AGGREGATION  METHOD FOR DETECTING MICROSATELLITE INSTABILITY FROM WHOLE-SLIDE IMAGES}
\name{\begin{tabular}{c}Kaifeng Pang $^{\star}$ $\quad$ Zuhayr Asad $^{\dagger}$ $\quad$ Shilin Zhao $^{\wr}$ $\quad$ Yuankai Huo $^{\dagger}$\end{tabular}}

\address{$^{\star}$ Nanjing University, School of Electronic Science and Engineering, China, 210023   \\
$^{\dagger}$ Vanderbilt University, Department of Computer Science, Nashville, TN, USA 37215  \\
$^{\wr}$ Vanderbilt University Medical Center, Department of Biostatistics, Nashville, TN, USA 37232 \\
(Corresponding author: Yuankai Huo)}

\usepackage{hyperref}

\begin{document}

\maketitle
\begin{abstract}
The prediction of microsatellite instability (MSI) and microsatellite stability (MSS) is essential in predicting both the treatment response and prognosis of gastrointestinal cancer. In clinical practice, a universal MSI testing is recommended, but the accessibility of such a test is limited. Thus, a more cost-efficient and broadly accessible tool is desired to cover the traditionally untested patients. In the past few years, deep-learning-based algorithms have been proposed to predict MSI directly from haematoxylin and eosin (H\&E)-stained whole-slide images (WSIs). Such algorithms can be summarized as (1) patch-level MSI/MSS prediction, and (2) patient-level aggregation. Compared with the advanced deep learning approaches that have been employed for the first stage, only the naïve first-order statistics (e.g., averaging and counting) were employed in the second stage. In this paper, we propose a simple yet broadly generalizable patient-level MSI aggregation (MAg) method to effectively integrate the precious patch-level information. Briefly, the entire probabilistic distribution in the first stage is modeled as histogram-based features to be fused as the final outcome with machine learning (e.g., SVM). The proposed MAg method can be easily used in a plug-and-play manner, which has been evaluated upon five broadly used deep neural networks: ResNet, MobileNetV2, EfficientNet, Dpn and ResNext. From the results, the proposed MAg method consistently improves the accuracy of patient-level aggregation for two publicly available datasets. It is our hope that the proposed method could potentially leverage the low-cost H\&E based MSI detection method. The code of our work has been made publicly available at \url{https://github.com/Calvin-Pang/MAg}.

\end{abstract}

\begin{figure*}[htbp]
\begin{minipage}[b]{1.0\linewidth}
  \centering
  \centerline{\includegraphics[width=13cm]{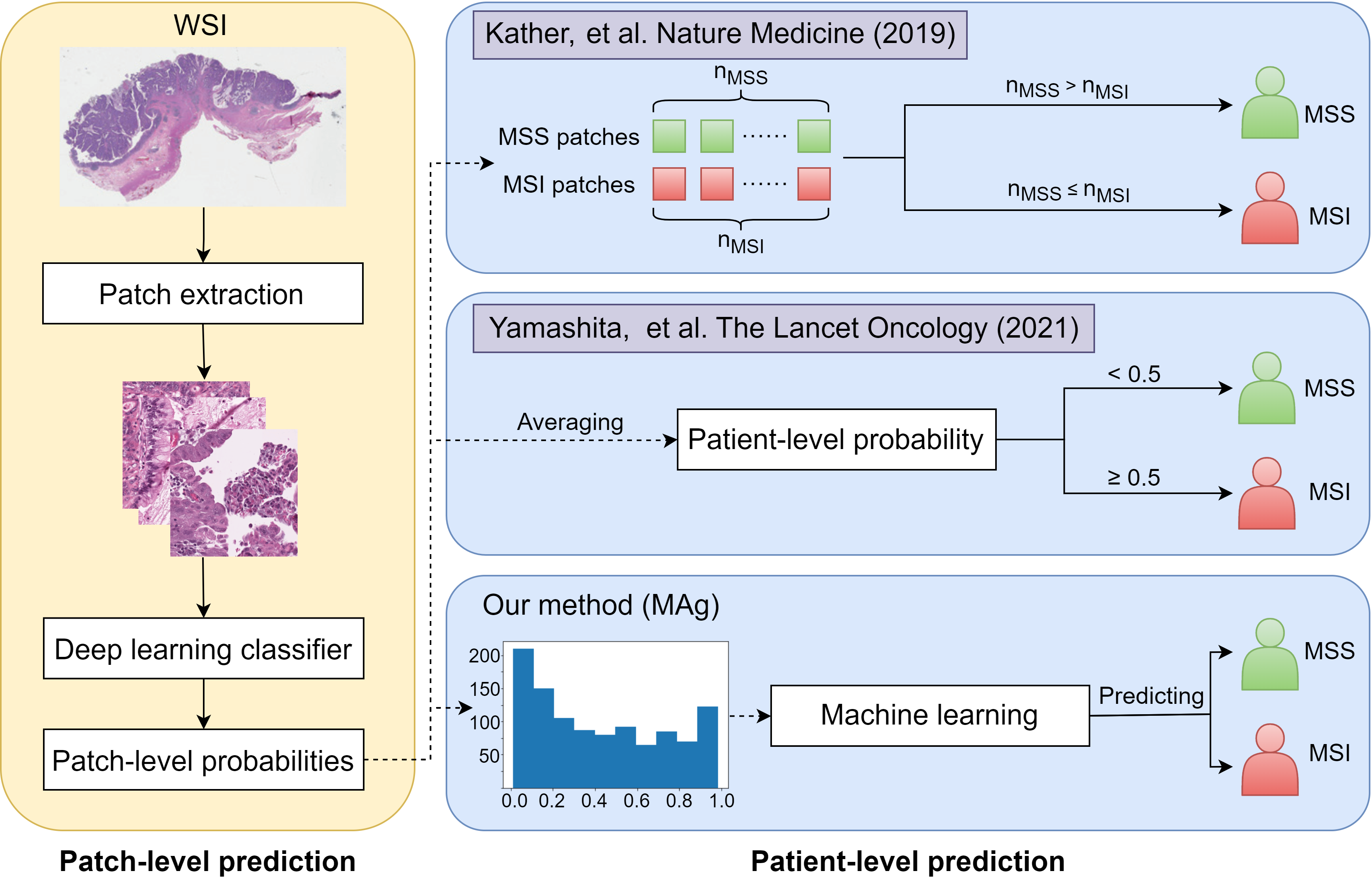}}
\end{minipage}
\caption{This figure shows the difference between the benchmarks and our proposed MAg method.} \vspace{-1em}
\label{fig1}
\end{figure*}

\vspace{-0.5em}
\section{Introduction}
\label{sec:intro}
The prediction of microsatellite instability (MSI) and microsatellite stability (MSS) plays an essential role in the diagnosis and treatment of gastrointestinal cancer~\cite{boland2010microsatellite}.  Unfortunately, the high-cost of the existing genetic test-based practice has been a primary hurdle of covering patients who need such a procedure. In the past few years, deep learning-based algorithms have shown promising results in directly predicting MSI and MSS from haematoxylin and eosin (H\&E)-stained whole-slide images (WSIs)~\cite{kather2019deep,yamashita2021deep}. Such methods can be summarized into two stages: \textbf{stage I}) image patch-level cancer prediction that computes the probability score of each patch, and \textbf{stage II}) patient-level characterization that aggregates the patch-level results into patient-level results. However, the major innovations and efforts focuses on \textbf{stage I} (e.g., deep learning approaches), while very few, if any, efforts are made for \textbf{stage II} beyond the first-order statistics methods (e.g., averaging and counting) as shown in \textbf{Fig. \ref{fig1}}. Specifically, Kather et al., simply calculate the proportion of patches judged to be MSI among all patches as the patient-level probability score~\cite{kather2019deep} (referred to as the “counting” baseline). Yamashita et al., directly calculate the average probability across all patches as the patient-level probability score~\cite{yamashita2021deep} (referred to as the “averaging” baseline). As a result, the precious patch-level predictions from deep learning approaches are not fully utilized beyond first-order statistics. We hypothesize that a more holistic consideration of patch-level probabilistic distribution using simple machine learning methods (e.g., SVM) could lead to consistently better patient-level performance of MSI detection.
As opposed to the previous work that focused on the innovations in \textbf{stage I}, we propose a simple but effective patient-level MSI aggregation (MAg) method by modeling overall probabilistic distribution with an SVM classifier. Briefly, in order to make full use of the precious patch-level information, we employ the histogram as a patient-level feature representation to model the distribution of probability scores from all patches within tumor regions. Then, a SVM classifier is utilized to further classify the histogram maps into two groups: MSI and MSS. 

The contribution of our research is three-fold: (1) we explore the previously ignored \textbf{stage II} patient-level MSI aggregation from a learning based perspective; (2) the proposed MAg method is a simple plug-and-play solution that is compatible with various \textbf{stage I} approaches; and (3) comprehensive evaluation has been performed upon five prevalent \textbf{stage I} approaches and two publicly available unique datasets. The experimental results demonstrate that our method consistently improve the F1 score and balanced accuracy (BACC) in its prediction of microsatellite instability (MSI vs. MSS) across different models and datasets.
\vspace{-1em}

\section{Method}
\label{sec:method}

\begin{figure*}[htbp]
\begin{minipage}[b]{1.0\linewidth}
  \centering
  \centerline{\includegraphics[width=16cm]{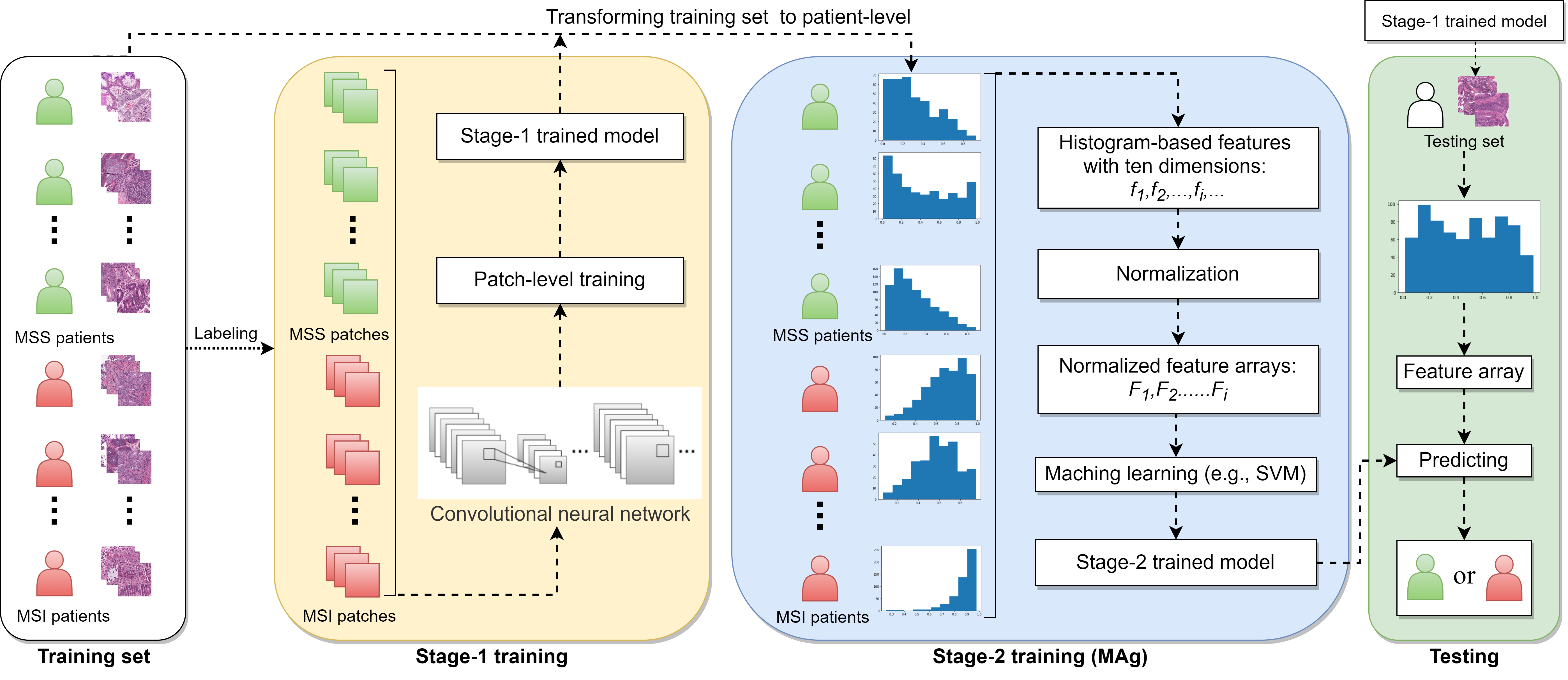}}
\end{minipage}
\caption{Shows the two-stage training and testing phases for the proposed method.} \vspace{-1em}
\label{fig2}
\end{figure*}

The proposed method (MAg) aggregates the overall patch-level probability scores into the patient-level scores by emphasizing the learning-based strategy in \textbf{stage II} (\textbf{Fig. \ref{fig2}}).

\vspace{-1em}

\subsection{Stage I: Patch-level classification}
\label{patch-level training}

In order to achieve the patch-level classification (\textbf{stage I}), ResNet~\cite{he2016deep} and MobileNetV2~\cite{sandler2018mobilenetv2} have been broadly employed in the prior arts~\cite{kather2019deep,yamashita2021deep}. In this paper, we further employ other prevalent classification benchmarks (i.e.,  Dpn131~\cite{chen2017dual}, EfficientNet~\cite{tan2019efficientnet}, ResNext~\cite{xie2017aggregated}) in order to evaluate the generalizability of the proposed plug-and-play MAg approach. For \textbf{stage I}, patients are divided into the same training, validation, and testing sets with the ratio of 50\%, 20\%, and 30\% respectively. The related image patches are extracted from WSIs following~\cite{kather2019deep,yamashita2021deep}. The experimental settings are presented in §4 as patch-level learning is not the major focus of this paper.

\vspace{-0.5em}

\subsection{Stage II: Patient-level aggregation with MAg}
\label{patient-level training}
As opposed to the prior arts~\cite{kather2019deep,yamashita2021deep} that simply compute first-order statistics (e.g., mean probability or class counting), we propose to model the overall distribution of patch-level probabilities as a patch-level histogram feature. Briefly, the histogram-based features describe the distribution features of the probability scores from all patches within the tumor region. Each histogram has ten bins as a ten-dimensional array that represents the histogram-based features of the patient: 
\begin{equation}
    f{_i} = [a{_i}{_0}, a{_i}{_1}, ... , a{_i}{_9}]
\end{equation}
where \emph{i} indicates different patients, and \emph{a$_0$} to \emph{a$_9$} represents the number of patches within each bin with 0.1 incremental probabilities. Moreover, in order to ensure that all the feature arrays have the same scale, a normalization is conducted to standardize the histograms for different patients:
\begin{equation}
    F{_i} = f{_i} / N{_i}
\end{equation}
where \emph{N} is the total number of patches for each patient.
After getting the histogram-based feature arrays of all patients, the original dataset is adapted from the patch-level results into patient-level features, with normalization  Eq.(2). Second, we employ a broadly-validated Support Vector Machine (SVM) that is used as the machine learning classifier given the sample size as well as the feature space. 

\subsection{Testing strategies}
\label{testing}
First, the trained classification models in \textbf{stage I} is used to achieve patch-level results to get the probability scores of all the patches in the test set. Second, we use the MAg method to aggregate the histogram-based features at patient-level. Then, we use the results in prediction and the ground truth labels at the patient-level to calculate the F1 score and balanced ACC (BACC) score.
\vspace{-1em}

\section{Data}
\label{sec:data}
 Two public datasets that include 411,890 patches derived from histological images of colorectal cancer and gastric cancer patients (n = 644) in the TCGA cohort were used and split into training, validation, and testing sets~\cite{kather2019histological}. All of the images in this repository were derived from formalin-fixed paraffin-embedded (FFPE) diagnostic slides. One dataset was acquired from the colorectal cancer TCGA patients (CRC dataset), while the other was acquired from the gastric (stomach) cancer TCGA patients (STAD dataset). In both datasets, patients were assigned with either ``MSS" (microsatellite stable) or ``MSIMUT" (microsatellite instable or highly mutated) labels. Each image patch was resized to 224 × 224 px at a resolution of 0.5 µm/px. A color normalization operation with the Macenko method~\cite{macenko2009method} was used for normalizing such datasets. This research study was conducted retrospectively using human subject data made available in open access by~\cite{kather2019histological}. Ethical approval was not required as confirmed by the license attached with the open access data. 

\begin{table}[h]
\caption{Results in the CRC dataset}
\centering
\begin{tabular}{c|c|ccc}
\hline
Model                                  & Metrics & \begin{tabular}[c]{@{}c@{}}Kather,\\ et al.~\cite{kather2019deep}\end{tabular} & \begin{tabular}[c]{@{}c@{}}Yamashita,\\ et al.~\cite{yamashita2021deep}\end{tabular} & \begin{tabular}[c]{@{}c@{}}MAg\\ (Ours)\end{tabular}             \\ \hline
\multirow{2}{*}{\textbf{ResNet}}       & F1      & 0.6032                                                   & 0.6032                                                      & \textbf{0.6230} \\ \cline{2-5} 
                                       & BACC    & 0.7438                                                   & 0.7438                                                      & \textbf{0.7573} \\ \hline
\multirow{2}{*}{\textbf{MobileNetV2}}  & F1      & 0.4324                                                   & 0.4324                                                      & \textbf{0.6122} \\ \cline{2-5} 
                                       & BACC    & 0.6336                                                   & 0.6336                                                      & \textbf{0.7344} \\ \hline
\multirow{2}{*}{\textbf{EfficientNet}} & F1      & 0.5902                                                   & 0.5806                                                      & \textbf{0.6154} \\ \cline{2-5} 
                                       & BACC    & 0.7313                                                   & 0.7245                                                      & \textbf{0.7562} \\ \hline
\multirow{2}{*}{\textbf{Dpn}}          & F1      & 0.5714                                                   & 0.5714                                                      & \textbf{0.6316} \\ \cline{2-5} 
                                       & BACC    & 0.7084                                                   & 0.7084                                                      & \textbf{0.7583} \\ \hline
\multirow{2}{*}{\textbf{ResNext}}      & F1      & 0.4528                                                   & 0.4615                                                      & \textbf{0.5574} \\ \cline{2-5} 
                                       & BACC    & 0.6294                                                   & 0.6362                                                      & \textbf{0.7053} \\ \hline
\end{tabular}\vspace{-0.8em}
\label{table1}
\end{table}

\begin{table}[h]
\caption{Results in the STAD dataset}
\centering
\begin{tabular}{c|c|ccc}
\hline
Model                                  & Metrics & \begin{tabular}[c]{@{}c@{}}Kather,\\ et al.~\cite{kather2019deep}\end{tabular} & \begin{tabular}[c]{@{}c@{}}Yamashita,\\ et al.~\cite{yamashita2021deep}\end{tabular} & \begin{tabular}[c]{@{}c@{}}MAg\\ (Ours)\end{tabular}             \\ \hline
\multirow{2}{*}{\textbf{ResNet}}       & F1      & 0.5517                                                   & 0.5614                                                      & \textbf{0.6038} \\ \cline{2-5} 
                                       & BACC    & 0.7051                                                   & 0.7119                                                      & \textbf{0.7389} \\ \hline
\multirow{2}{*}{\textbf{MobileNetV2}}  & F1      & 0.5763                                                   & 0.5614                                                      & \textbf{0.5862} \\ \cline{2-5} 
                                       & BACC    & 0.7251                                                   & 0.7119                                                      & \textbf{0.7319} \\ \hline
\multirow{2}{*}{\textbf{EfficientNet}} & F1      & 0.5763                                                   & 0.5763                                                      & 0.5763          \\ \cline{2-5} 
                                       & BACC    & 0.7251                                                   & 0.7251                                                      & 0.7251          \\ \hline
\multirow{2}{*}{\textbf{Dpn}}          & F1      & 0.5000                                                   & 0.4906                                                      & \textbf{0.5185} \\ \cline{2-5} 
                                       & BACC    & 0.6651                                                   & 0.6586                                                      & \textbf{0.6786} \\ \hline
\multirow{2}{*}{\textbf{ResNext}}      & F1      & 0.4528                                                   & 0.4528                                                      & \textbf{0.5231} \\ \cline{2-5} 
                                       & BACC    & 0.6319                                                   & 0.6319                                                      & \textbf{0.6846} \\ \hline
\end{tabular}\vspace{-0.8em}
\label{table2}
\end{table}

\section{experiments and results}
\label{experiments}
The experiments were performed on a Google Colab workstation with a NVIDIA Tesla P100 GPU.

\textbf{Stage I.} All the patch-wise classification benchmarks were pretrained by ImageNet~\cite{russakovsky2015imagenet}. In this stage, we set the learning ratio as 0.0001. The number of epochs was set at 40 with a mini-batch size of 32. A binary cross entropy loss function and the Adam optimizer  were applied. We augmented the training data by randomly introducing positional transforms and color transforms. Finally, we used the accuracy of the validation set to select the best trained model from the 40 epochs. In order to verify the model universality of our proposed MAg method, five commonly used deep learning classification models were used in patch level, and their fully connected layers were modified according to the output dimension. The corresponding low-dimentional representation features are the results of \textbf{stage I}.

In \textbf{stage I}, five prevalent approaches have been used to be the baseline feature extractors, including \textbf{ResNet}~\cite{he2016deep}, \textbf{MobileNetV2}~\cite{sandler2018mobilenetv2}, \textbf{EfficientNet}~\cite{tan2019efficientnet}, \textbf{Dpn}~\cite{chen2017dual}, and  \textbf{ResNext}~\cite{xie2017aggregated} models.







\textbf{Stage II.} The second stage performed the proposed MAg method at patient-level, as well as comparing it with the baseline counting method~\cite{kather2019deep} and the averaging method~\cite{yamashita2021deep}. In the experiments, we employed the same patient-level threshold = 0.5, that is, patients with a final probability score greater than or equal to 0.5 were judged as MSI, and patients with a final probability score less than 0.5 were judged as MSS. Moreover, to assess the generalizability, the experiments above were done in both the CRC dataset and the STAD dataset.

\textbf{Table. \ref{table1}} shows the results of the three patient-level aggregation methods via five different patch-level training models on the CRC dataset. \textbf{Table. \ref{table2}} shows the the comparison of the results of the three patient-level aggregation methods after using five different patch-level training models on the STAD dataset.

From the quantitative results, the proposed simple MAg strategy achieved higher F1 scores and balanced accuracy scores than the two commonly used methods, the counting method and the averaging method, on different models and the two unique datasets. It demonstrated that the MAg method is a generalizable \textbf{stage II} approach that consistently improved the performance upon different \textbf{stage I} models.

\section{Conclusions}
\label{sec:conclusion}
In this paper, we propose a histogram-based MSI aggregation (MAg) method in the prediction of  microsatellite instability. The experimental results show that the histogram-based features of the patch-level probabilities can effectively improve the classification ability of the previous commonly used methods. Moreover, such \textbf{stage II} improvements are consistent across different \textbf{stage I} methods. We hope this pipeline can become a potential routine operation for optimizing the classification model in microsatellite instability prediction. \\
\vspace{-1.5em}

\section{Compliance with Ethical Standard}
\label{sec:ethic}
This research study was conducted retrospectively using human subject data made available in open access by~\cite{kather2019histological}. Ethical approval was not required as confirmed by the license attached with the open access data. There is no conflicts of interests of all authors.

\vspace{-0.5em}
\bibliographystyle{IEEEbib}
\bibliography{reference_K}

\begin{thebibliography}{10}
\providecommand{\url}[1]{#1}
\csname url@samestyle\endcsname
\providecommand{\newblock}{\relax}
\providecommand{\bibinfo}[2]{#2}
\providecommand{\BIBentrySTDinterwordspacing}{\spaceskip=0pt\relax}
\providecommand{\BIBentryALTinterwordstretchfactor}{4}
\providecommand{\BIBentryALTinterwordspacing}{\spaceskip=\fontdimen2\font plus
\BIBentryALTinterwordstretchfactor\fontdimen3\font minus
  \fontdimen4\font\relax}
\providecommand{\BIBforeignlanguage}[2]{{%
\expandafter\ifx\csname l@#1\endcsname\relax
\typeout{** WARNING: IEEEtran.bst: No hyphenation pattern has been}%
\typeout{** loaded for the language `#1'. Using the pattern for}%
\typeout{** the default language instead.}%
\else
\language=\csname l@#1\endcsname
\fi
#2}}
\providecommand{\BIBdecl}{\relax}
\BIBdecl

\bibitem{boland2010microsatellite}
C.~R. Boland and A.~Goel, ``Microsatellite instability in colorectal cancer,''
  \emph{Gastroenterology}, vol. 138, no.~6, pp. 2073--2087, 2010.

\bibitem{kather2019deep}
J.~N. Kather, A.~T. Pearson, N.~Halama, D.~J{\"a}ger, J.~Krause, S.~H. Loosen,
  A.~Marx, P.~Boor, F.~Tacke, U.~P. Neumann \emph{et~al.}, ``Deep learning can
  predict microsatellite instability directly from histology in
  gastrointestinal cancer,'' \emph{Nature medicine}, vol.~25, no.~7, pp.
  1054--1056, 2019.

\bibitem{yamashita2021deep}
R.~Yamashita, J.~Long, T.~Longacre, L.~Peng, G.~Berry, B.~Martin, J.~Higgins,
  D.~L. Rubin, and J.~Shen, ``Deep learning model for the prediction of
  microsatellite instability in colorectal cancer: a diagnostic study,''
  \emph{The Lancet Oncology}, vol.~22, no.~1, pp. 132--141, 2021.

\bibitem{russakovsky2015imagenet}
O.~Russakovsky, J.~Deng, H.~Su, J.~Krause, S.~Satheesh, S.~Ma, Z.~Huang,
  A.~Karpathy, A.~Khosla, M.~Bernstein \emph{et~al.}, ``Imagenet large scale
  visual recognition challenge,'' \emph{International journal of computer
  vision}, vol. 115, no.~3, pp. 211--252, 2015.

\bibitem{he2016deep}
K.~He, X.~Zhang, S.~Ren, and J.~Sun, ``Deep residual learning for image
  recognition,'' in \emph{Proceedings of the IEEE conference on computer vision
  and pattern recognition}, 2016, pp. 770--778.

\bibitem{sandler2018mobilenetv2}
M.~Sandler, A.~Howard, M.~Zhu, A.~Zhmoginov, and L.-C. Chen, ``Mobilenetv2:
  Inverted residuals and linear bottlenecks,'' in \emph{Proceedings of the IEEE
  conference on computer vision and pattern recognition}, 2018, pp. 4510--4520.

\bibitem{chen2017dual}
Y.~Chen, J.~Li, H.~Xiao, X.~Jin, S.~Yan, and J.~Feng, ``Dual path networks,''
  \emph{arXiv preprint arXiv:1707.01629}, 2017.

\bibitem{tan2019efficientnet}
M.~Tan and Q.~Le, ``Efficientnet: Rethinking model scaling for convolutional
  neural networks,'' in \emph{International Conference on Machine
  Learning}.\hskip 1em plus 0.5em minus 0.4em\relax PMLR, 2019, pp. 6105--6114.

\bibitem{xie2017aggregated}
S.~Xie, R.~Girshick, P.~Doll{\'a}r, Z.~Tu, and K.~He, ``Aggregated residual
  transformations for deep neural networks,'' in \emph{Proceedings of the IEEE
  conference on computer vision and pattern recognition}, 2017, pp. 1492--1500.

\bibitem{kather2019histological}
J.~Kather, ``Histological images for msi vs. mss classification in
  gastrointestinal cancer, ffpe samples,'' \emph{ZENODO}, 2019.

\bibitem{macenko2009method}
M.~Macenko, M.~Niethammer, J.~S. Marron, D.~Borland, J.~T. Woosley, X.~Guan,
  C.~Schmitt, and N.~E. Thomas, ``A method for normalizing histology slides for
  quantitative analysis,'' in \emph{2009 IEEE International Symposium on
  Biomedical Imaging: From Nano to Macro}.\hskip 1em plus 0.5em minus
  0.4em\relax IEEE, 2009, pp. 1107--1110.

\end{thebibliography}

\end{document}